\documentclass[12pt,prd,preprint,nofootinbib]{revtex4}
\usepackage{amssymb}
\usepackage{bm}
\usepackage{enumerate}
\usepackage{amssymb}
\usepackage{amsmath}
\usepackage{feynmf}
\usepackage{slashed}
\usepackage{graphicx}

\newcommand\aea{&=&}
\newcommand\be{\begin{equation}}
\newcommand\ee{\end{equation}}

\newcommand\bea{\begin{eqnarray}}
\newcommand\eea{\end{eqnarray}}

\begin{document}

\title{Stability Aspects of Wormholes in $R^2$ Gravity\\ \vspace{1cm}}
\author{James B. Dent$^{\bf a}$}
\author{Damien A. Easson$^{\bf b}$}
\author{Thomas W. Kephart$^{\bf c}$}
\author{Sara C. White$^{\bf a}$}
\affiliation{$^{\bf a}$ Department of Physics, University of Louisiana at Lafayette, \\
Lafayette, LA 70504, USA\\}
\affiliation{$^{\bf b}$ Department of Physics    \& Beyond Center for Fundamental Concepts in Science,\\
Arizona State University, Tempe, AZ 85287-1504, USA \\}
\affiliation{$^{\bf c}$ Department of Physics and Astronomy, Vanderbilt University, Nashville, TN 37235, USA \\ \vspace{2cm}
}
\vspace{2cm}

\begin{abstract}
We study radial perturbations of a wormhole in $R^2$-gravity to determine regions of stability. We also investigate  
massive and massless particle orbits and tidal forces in this space-time for a radially infalling observer.
\end{abstract}

\maketitle

\section{Introduction}


Wormholes are intriguing solutions to Einstein's theory of  general relativity, though their physical existence is highly contested. Wormholes are shortcut paths between different parts of spacetime and have long served as a theoretical laboratory for exploring exotic phenomena in general relativity \cite{Morris:1988tu,Morris:1988cz,Visser:1995cc}. 
Wormhole solutions are found by specifying the desired spacetime through sewing together two separate spacetimes with black hole geometries. This procedure determines the type of energy-momentum tensor necessary to support such a solution and typically describes  ``exotic" types of mass and energy.  There are longstanding discussions regarding the viability of constructing wormhole solutions in standard general relativity. 
Surprisingly, wormholes in a certain modification of general relativity, namely $R = 0$ solutions of $R^2$ gravity,  exist without the need for exotic matter  \cite{Duplessis:2015xva}.  $R^2$ gravity has recently gained attention due to the discovery 
of new spherically symmetric solutions in the $R=0$ regime \cite{Kehagias:2015ata}, including  black hole solutions in general quadratic gravity \cite{Lu:2015cqa}.  One of the intriguing features of
$R^2$ gravity is, unlike in General Relativity, it is possible to have non-trivial Ricci tensor $R_{\mu \nu} \neq 0$ even when $R=0$.

In this paper we study stability properties of the aforementioned $R^2$ gravity wormhole solutions. We identify regions in parameter space which yield a stable wormhole by studying radial perturbations. In addition, we search for stable orbits around these wormholes and determine whether an astrophysical probe in an $R^2$ universe could possibly traverse such a wormhole without experiencing destructive tidal forces. The present work explores these questions by adapting previous stability inquiries for wormholes in standard general relativity such as those discussed in \cite{Poisson:1995sv}.

We begin in section \ref{Wormhole Construction} by reviewing the wormhole construction found in \cite{Duplessis:2015xva}, and in section \ref{$f(R)$ gravity junction conditions} we determine the junction conditions needed in $f(R)$ gravity. In section \ref{Stability} we discuss the criteria for a stable wormhole and find regions in parameter space where stability holds. Section \ref{Orbits} examines the stability of orbits around the stable wormhole construction, and section \ref{Tidal Forces} determines the tidal forces encountered by a radially infalling observer passing through the wormhole. We conclude in section \ref{Conclusion}.

\section{Wormhole Construction} \label{Wormhole Construction}

 Consider the wormhole $R=0$ solution to $R^2$ gravity. $R=0$ is a requirement 
 for this vacuum solution. 
 The metric is
\bea\label{metric}
ds^2 = -G(\ell)dt^2 + \frac{d\ell^2}{G(\ell)} + (\ell^2+k^2)d\Omega^2 \,
\eea
where we define
\bea
\ell^2 = r^2 - k^2 .
\eea
Here, $l$ is the radius of the wormhole throat and $d\Omega^2$ is the metric on the unit two-sphere. The constant $k$ sets the minimal throat radius, and the full form of $G(\ell)$ can be found in the Appendix of \cite{Duplessis:2015xva}. If $G(\ell) \rightarrow 1$, then $\ell\rightarrow \pm \infty $ and the space is asymptotically Minkowski. For large $\ell$ we have $G(\ell) \rightarrow 1-\frac{2M_{\pm}}{\ell}+\frac{Q^2_{\pm}}{\ell}$ and the space is similar to Reissner-Nordstr\"om. (We shall not need the definitions of $M_{\pm}$ and $Q^2_{\pm}$, but they can be found in \cite{Duplessis:2015xva}.) Therefore, allowing the throat at $r=a$ to have time dependence, we have the relations
\bea
\frac{d\ell}{d\tau} \aea \frac{a}{\ell}\frac{da}{d\tau}=\frac{a\dot{a}}{\ell}\\
\frac{d^2\ell}{d\tau^2} \aea \frac{a\ddot{a}}{\ell} - \frac{\dot{a}^2k^2}{\ell^3}\\
\frac{dG(\ell)}{d\tau} \aea \frac{dG}{d\ell}\frac{a\dot{a}}{\ell} = G'\frac{a\dot{a}}{\ell}
\eea
From equation (\ref{metric}) we find
\bea
\frac{dt}{d\tau} = \frac{\sqrt{G(\ell)\ell^2 + (\ell^2 + k^2)\dot{a}^2}}{G(\ell)\ell} = \frac{\sqrt{G+\dot{\ell}^2}}{G}
\eea
along with the timelike Killing vector
\bea
X_{\mu} = (-G(\ell),0,0,0)
\eea
and the four-velocity
\bea
U^{\mu} = (\frac{dt}{d\tau},\dot{\ell},0,0)
\eea
The extrinsic curvature $K^{\tau}{}_{\tau}$ is then given by
\bea
\frac{D}{D\tau}(X_{\mu}U^{\mu}) = -A\dot{\ell}
\eea
One can perform the $\tau$ derivative, which gives
\bea
K^{\tau}{}_{\tau} = \frac{\ddot{\ell} + G'(\ell)/2}{\sqrt{G(\ell) + \dot{\ell}^2}}
\eea
It can be seen that this is equivalent to using the replacements
\bea
F(r) &\rightarrow& G(\ell)\\
\dot{a} &\rightarrow& \dot{\ell}\\
\ddot{a} &\rightarrow& \ddot{\ell}
\eea
in the $K^{\tau}_{\tau}$ found from the metric
\bea\label{metric2}
ds^2 = -F(r)dt^2 + \frac{dr^2}{F(r)} + r^2d\Omega^2
\eea
This makes sense when directly comparing the metrics given in equations (\ref{metric}) and (\ref{metric2}).  Similar replacements can be made to recover $K^{\theta}{}_{\theta}$ and $K^{\phi}{}_{\phi}$.  We find
\bea
K^{\theta}{}_{\theta} = K^{\phi}{}_{\phi}=\frac{\sqrt{G+ \dot{\ell}^2}}{\ell}
\eea

\section{$f(R)$ gravity junction conditions} \label{$f(R)$ gravity junction conditions}

We have the energy-momentum on the throat given in terms of the extrinsic curvature
\bea
8\pi GS_{ij} = -\kappa_{ij} + h_{ij}\kappa
\eea
where
\bea
8\pi G S_{ij} = \frac{f-Rf'(R)}{2}h_{ij} + \nabla_i\nabla_jf'(R) - \Box f'(R)h_{ij}
\eea
and $\kappa_{ij}$ and $\kappa$ are given by the jump discontinuity equations for the extrinsic curvature
\bea
\kappa_{ij} \aea K_{ij}^+ - K_{ij}^-\\
\kappa \aea K^+ - K^-
\eea
with $K$ being the trace of the extrinsic curvature
\bea
K = K^{\tau}{}_{\tau} + K^{\theta}{}_{\theta} + K^{\phi}{}_{\phi}
\eea
In the case of our metric (\ref{metric}) we have the following relations of importance
\bea
R_{tt} \aea \frac{1}{2}G\left[\frac{2\ell G'}{\ell^2+k^2} + G''\right]\\
R_{rr} \aea -\frac{1}{2G}\left[\frac{4k^2G+2\ell(\ell^2+k^2)G'}{(\ell^2+k^2)^2} + G''\right]\\
R_{\theta\theta} \aea 1-G-\ell G'\\
R_{\phi\phi} \aea R_{\theta\theta}\textrm{sin}^2\theta \\
R \aea -\frac{(\ell^2+k^2)[(\ell^2+k^2)G'' + 4\ell G' - 2] + 2(\ell^2+2k^2)G}{(\ell^2+k^2)^2}\\
f(R) \aea R^2 \Longrightarrow f'(R) = 2R\\
K \aea \frac{\ddot{\ell}\ell + \frac{1}{2}G'\ell + 2G + 2\dot{\ell}^2}{\ell\sqrt{G + \dot{\ell}^2}}
\eea

\section{Stability} \label{Stability}

Apparently $f(R)$ models are all subject to the junction condition $[K] = 0$ \cite{Eiroa:2015hrt} in order for there to be no jump discontinuity between the two spaces. In our case this becomes
\bea
\ddot{\ell} = - \frac{G'(\ell)}{2} - \frac{2}{\ell}(G(\ell) + \dot{\ell}^2)
\eea
where we will be interested in the static case where $\dot{\ell} = 0$.  This leads to
\bea
G'(\ell_0) = -4\frac{G(\ell_0)}{\ell_0}
\eea
Defining the potential
\bea
V(\ell) = -\dot{\ell}^2
\eea
we find
\bea
V''(\ell_0) = G''(\ell_0) - \frac{20}{\ell_0^2}G(\ell_0)
\eea
Stability is then given by the condition $V''(\ell_0) > 0$.

\begin{figure}[h]
\includegraphics[scale = .8]{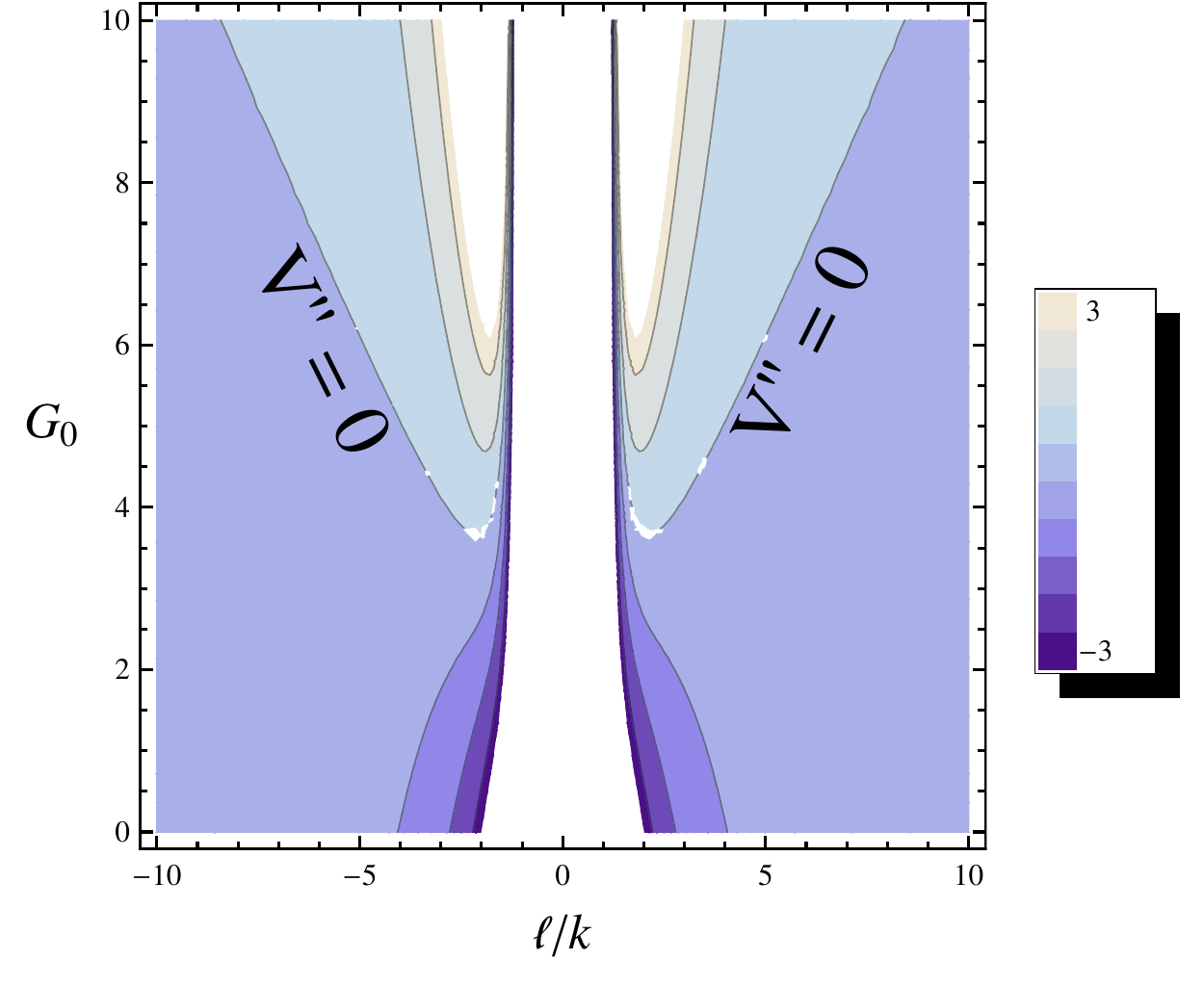}%
\caption{{\label{Plot1Jan14.pdf}Contour plot of $V''$ varying $G_0$ and  $\ell/k$ for the parameter values $G'(\ell) = 1.0$ and $k=1.0$.}}
\end{figure}

\begin{figure}[h]
\includegraphics[scale = .8]{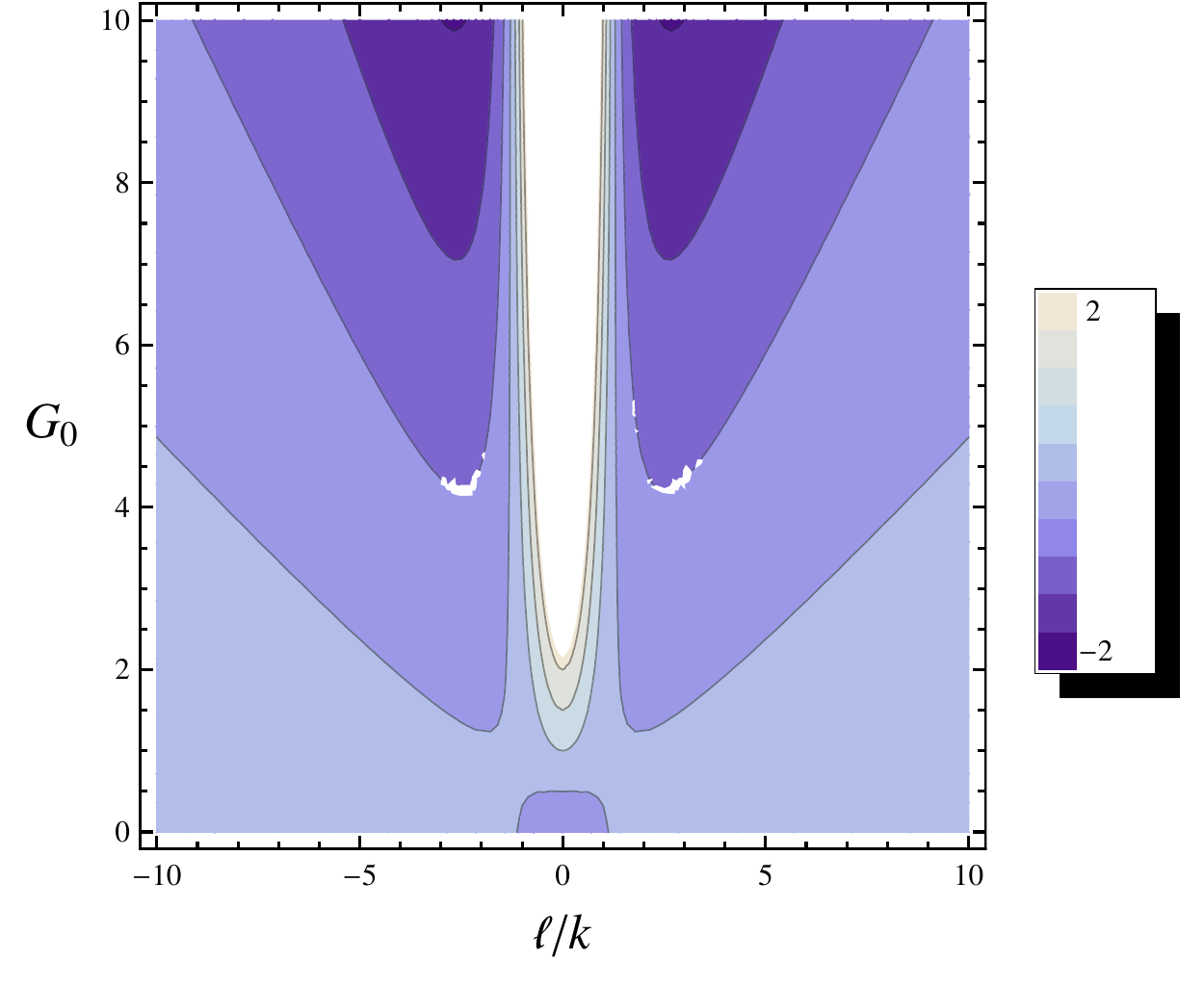}%
\caption{{\label{Plot2Jan14.pdf}Contour plot of $G$ varying $G_0$ and $\ell/k$ for the parameter values $G'(\ell) = 1.0$ and $k=1.0$.}}
\end{figure}

\begin{figure}[h]
\includegraphics[scale = .8]{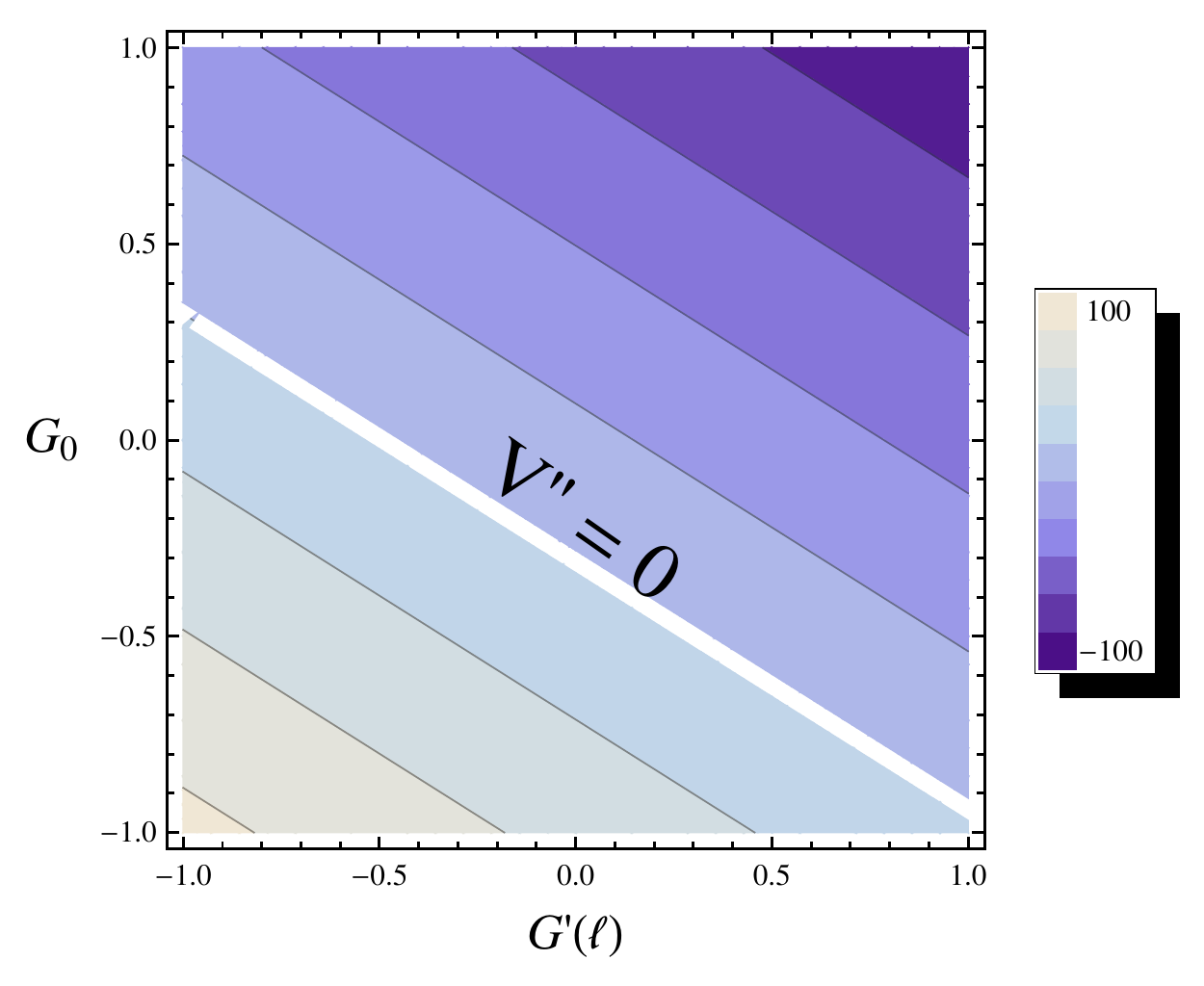}%
\caption{{\label{Plot3Jan14.pdf}Contour plot of $V''$ varying $G_0$ and $G'(\ell)$ for the parameter values $\ell/k = 0.5$ and $k = 1.0$.}}

\includegraphics[scale = .8]{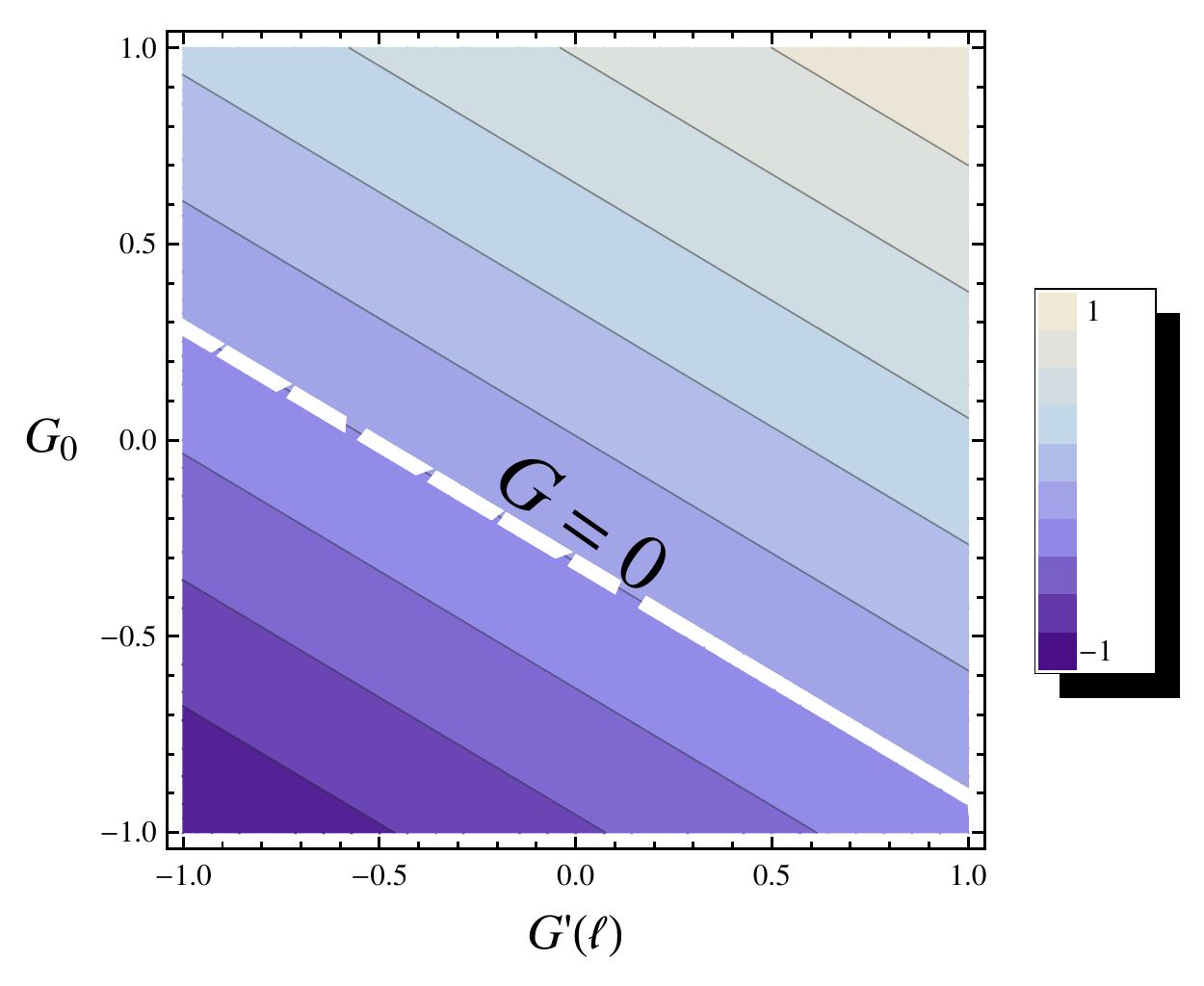}%
\caption{{\label{Plot4Jan14.pdf}Contour plot of $G$ varying $G_0$ and $G'(\ell)$ for the parameter values $\ell/k = 0.5$ and $k=1.0$.}}
\end{figure}

We can see in Figures (\ref{Plot1Jan14.pdf}-\ref{Plot4Jan14.pdf}) that  regions of stability exist for certain parameters.

\section{Orbits around the wormhole} \label{Orbits}

Beginning with the explicit form of the metric
\bea
ds^2 = -G(\ell)dt^2 + \frac{1}{G(\ell)}d\ell^2 + (\ell^2 + k^2)d\theta^2 + (\ell^2 + k^2)\textrm{sin}^2\theta d\phi^2
\eea
we examine equatorial orbits with $\theta = \pi/2$, where we have the geodesic equations
\bea
0 \aea \frac{d^2\ell}{d\lambda^2} - \frac{G'(\ell)}{2G(\ell)}\left(\frac{d\ell}{d\lambda}\right)^2 + \frac{G'(\ell)G(\ell)}{2}\left(\frac{dt}{d\lambda}\right)^2 - \ell G(\ell)\left(\frac{d\phi}{d\lambda}\right)^2\\
0 \aea \frac{d^2\phi}{d\lambda^2} + \frac{2\ell}{\ell^2 + k^2} \frac{d\phi}{d\lambda}\frac{d\ell}{d\lambda}\\
0 \aea \frac{d^2t}{d\lambda^2} + \frac{G'(\ell)}{G(\ell)}\frac{d\ell}{d\lambda}\frac{dt}{d\lambda}
\eea
The $\phi$ equation can be written as
\bea
\frac{1}{\ell^2+k^2}\frac{d}{d\lambda}\left((\ell^2+k^2)\frac{d\phi}{d\lambda}\right) = 0
\eea
Which leads to the constant of motion
\bea
(\ell^2+k^2)\frac{d\phi}{d\lambda} &\equiv& L
\eea
The $t$ equation can be written as
\bea
\frac{d}{d\lambda}\left(G(\ell)\frac{dt}{d\lambda}\right) = 0
\eea
This gives the constant
\bea
G(\ell)\frac{dt}{d\lambda} = E
\eea
We also know that the quantity
\bea
g_{\mu\nu}\frac{dx^{\mu}}{d\lambda}\frac{dx^{\nu}}{d\lambda}
\eea
is a constant along the path parameterized by $\lambda$.  If $d\lambda = d\tau$ then this is $-1$ for a massive particle, and it is always equal to zero for a massless particle.  Following \cite{Carroll:1997ar} we will call this constant $-\epsilon$, which then gives
\bea
-\epsilon \aea -G(\ell)\left(\frac{dt}{d\lambda}\right)^2 + \frac{1}{G(\ell)}\left(\frac{d\ell}{d\lambda}\right)^2 + (\ell^2 +k^2)\left(\frac{d\phi}{d\lambda}\right)^2\\
\aea -\frac{E^2}{G(\ell)} + \frac{1}{G(\ell)}\left(\frac{d\ell}{d\lambda}\right)^2 + \frac{L^2}{\ell^2 +k^2}
\eea

This can be written as
\bea
\frac{1}{2}\left(\frac{d\ell}{d\lambda}\right)^2 + V_{eff}(\ell) = \frac{E^2}{2}
\eea
with the effective potential defined as
\bea
V_{eff}(\ell) = \frac{1}{2}\left(\epsilon G(\ell) + \frac{G(\ell)L^2}{\ell^2+k^2}\right)
\eea
One can then find the stable orbits for $V_{eff}' = 0$ and $V_{eff}'' > 0$.

\begin{figure}[h]
\includegraphics[scale = 1.0]{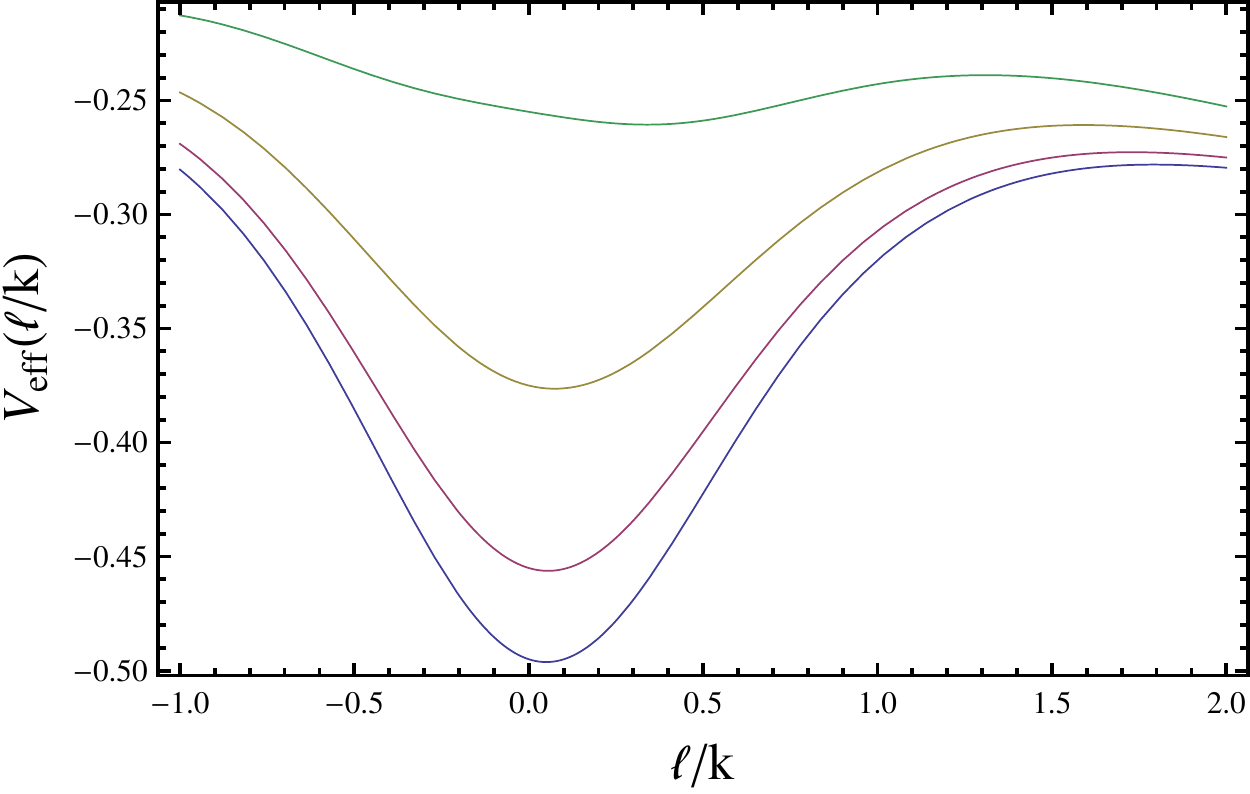}
\caption{\label{orbitplot1.pdf} A plot of $V_{eff}(x)$ as a function of $x \equiv \ell/k$ with $L =  0.1, 0.3, 0.5, 0.7$ which are the blue, red, gold, and green curves, respectively, and $k = 1.0$, $G_0 = 1.0$, and $v_0 = 0.1$.  }
\end{figure}

Figure \ref{orbitplot1.pdf} demonstrates there exists a region of orbital stability due to the presence of a minimum for the effective potential, and it is \emph{just} on the positive $x$ side as $L$ is lowered towards zero ($x = .05$ is roughly where the minimum value occurs as $L \rightarrow 0$). Unstable orbits are seen to exist when maxima of the effective potential are present. 

\section{Tidal Forces} \label{Tidal Forces}

By looking at the difference in acceleration between two neighboring points we are able to determine the tidal forces that a radially infalling observer will experience. To do this we use the equation for the tidal acceleration previously found in \cite{Duplessis:2015xva} given as
\bea
\Delta a^{\hat{j}} = -R^{\hat{j}}_{\hat{\alpha}\hat{\beta}\hat{\rho}}u^{\hat{\alpha}}\xi^{\hat{\beta}}u^{\hat{\rho}} = -R_{\hat{j}\hat{0}\hat{k}\hat{0}}\xi^{\hat{k}}
\eea
where $\xi$ is the separation distance between the two points, measured by our infalling observer, and $u$ is the four-velocity. 

The Riemann tensors, which determine the tidal forces for the metric (\ref{metric}) are found in \cite{Duplessis:2015xva}, and here we reproduce them in terms of the variable $x$
\bea
&&R_{\hat{1}\hat{0}\hat{1}\hat{0}} = -\frac{G''}{2k^2}\\
&&R_{\hat{2}\hat{0}\hat{2}\hat{0}} = R_{\hat{3}\hat{0}\hat{3}\hat{0}} = \gamma^2\frac{xG'}{k^2(x^2+1)} - v^2\gamma^2\left(\frac{k^2(x^2+1)G' + 2k^2G}{2k^4(x^2+1)^2}\right)
\eea
In order to determine if the infalling observer will survive the journey through the wormhole, we now examine the tidal forces in the range of parameter values previously found to produce a stable wormhole configuration. As a benchmark, and following the traditional treatment found in \cite{Morris:1988cz}, we will utilize the acceleration standard of the gravitational acceleration, $g$ (9.8m/s$^2$ in SI units), at the surface of the Earth.  

\begin{figure}[h]
\includegraphics[scale = 1.0]{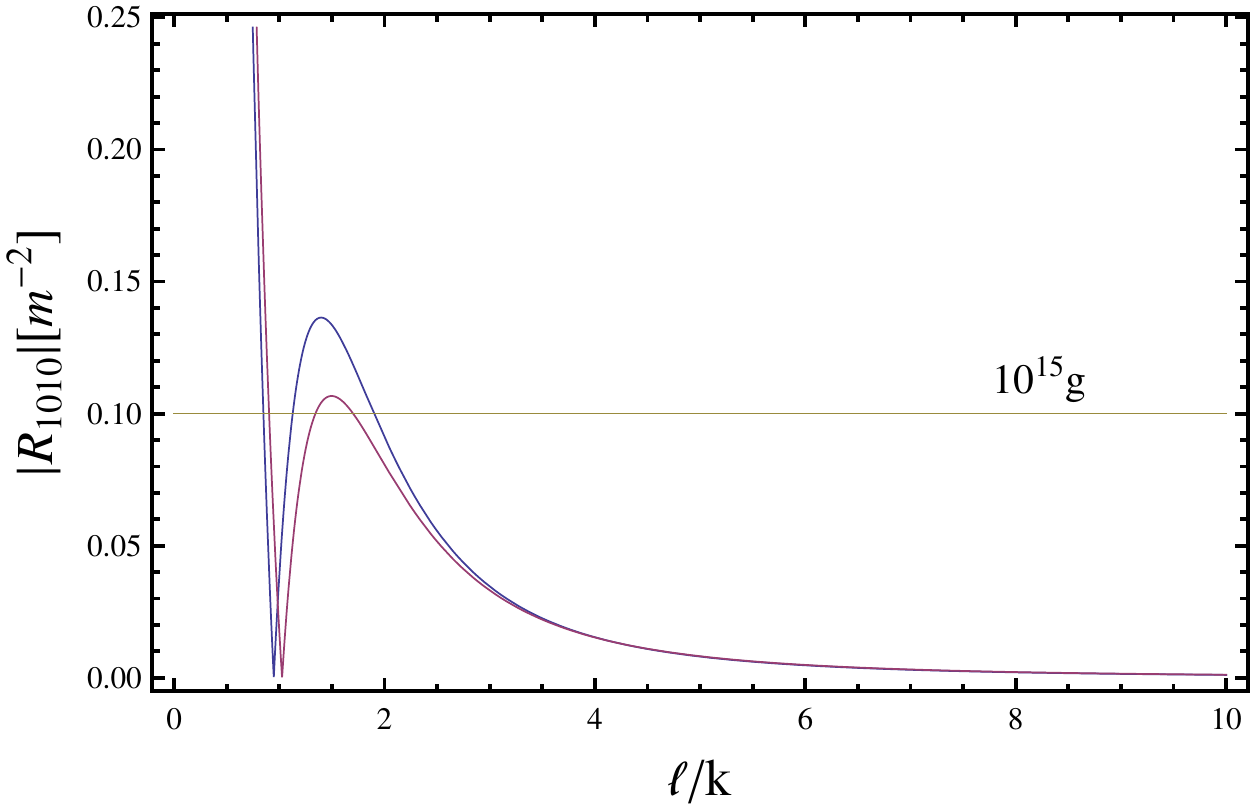}
\caption{\label{tidal1.pdf} This is a plot of the absolute value of $R_{1010}$ for the parameter values $k = 1.0$, $G_0 = 0.1$ (blue), $G_0 = 0.2$ (red), and $v_0 = -0.9$.}
\end{figure}

\begin{figure}[h]
\includegraphics[scale = 1.0]{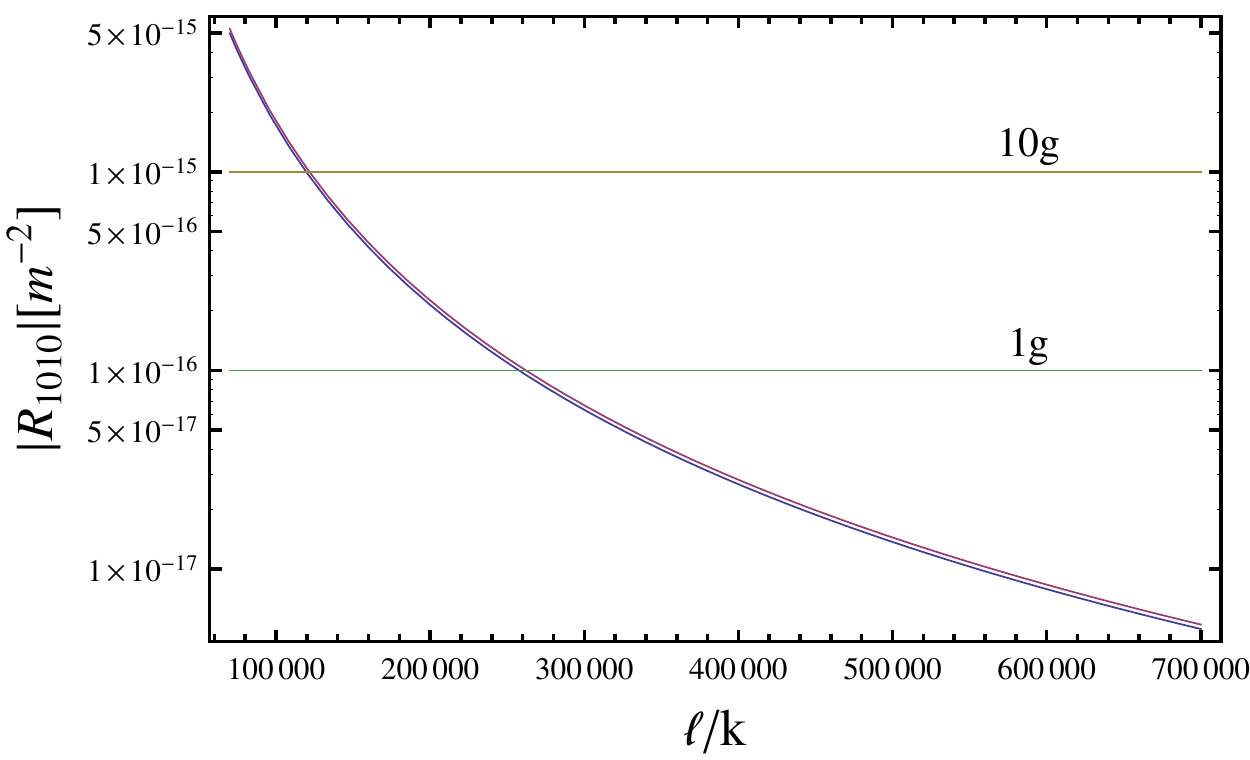}
\caption{\label{accelplot1.pdf} This is a plot of the absolute value of $R_{1010}$ for the parameter values $k = 1.0$, $G_0 = 0.1$ (blue), $G_0 = 0.2$ (red), and $v_0 = -0.9$.}

\includegraphics[scale = 1.0]{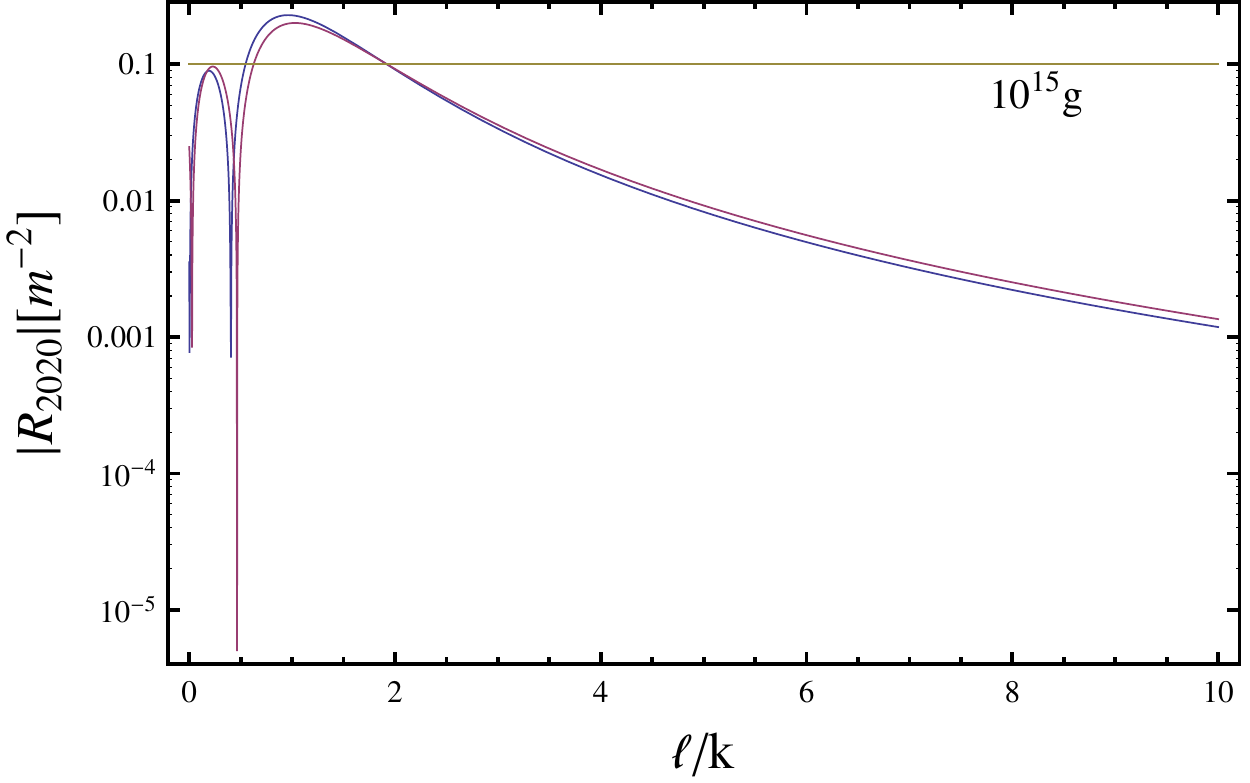}
\caption{\label{R2020plot1.pdf} This is a plot of the absolute value of $R_{2020}$ for the parameter values $k = 1.0$, $G_0 = 0.1$ (blue), $G_0 = 0.2$ (red), and $v_0 = -0.9$.}
\end{figure}

\begin{figure}[h]
\includegraphics[scale = 1.0]{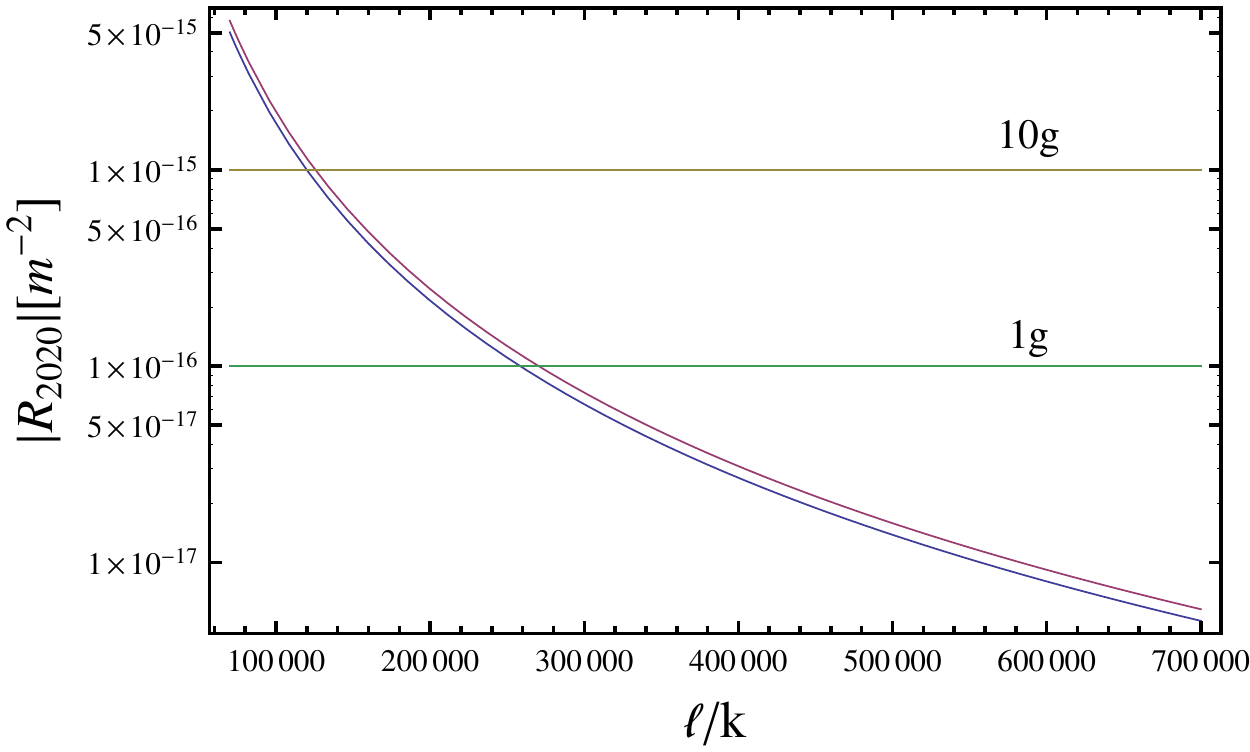}
\caption{\label{R2020plot2.pdf} This is a plot of the absolute value of $R_{2020}$ for the parameter values $k = 1.0$, $G_0 = 0.1$ (blue), $G_0 = 0.2$ (red), and $v_0 = -0.9$.}
\end{figure}

The absolute values of the Riemann tensors are translated to factors of $g$ as shown in figures (\ref{tidal1.pdf}-\ref{R2020plot2.pdf}). In these figures, the parameter $k$ set to $1.0$ meter implies that the $\ell/k$ values give $\ell$ in meters.  Unsurprisingly, for very small $\ell$ values, equivalently small wormhole throats, the tidal forces produced are immense, but decline precipitously with increasing $\ell$, reaching values sustainable by humans (or human-built spacecraft) at only a few hundred km for these parameter values.

\section{Discussion and Conclusion} \label{Conclusion}

While $R^2$ gravity may not correspond to physical reality, it provides a testing ground where we can find analytic solutions
and study their corresponding stability. (The case has been made for the physical significance of conformal gravity \cite{Mannheim:2011ds}, which is also quadratic in the curvature.) 
Hence the solutions and properties thereof discussed here may be relevant to more physically well motivated general theories of quadratic gravity.
Here we have found stability conditions for radially perturbed wormholes in $R^2$ gravity with $R=0$ that do not require exotic matter.
Properties of stable particle orbits in these solutions then give us hints of what we may expect to find in classes of more general theories that are quadratic in curvature.

\section{Acknowledgements}
 J.B.D. thanks Dr. and Mrs. Sammie W. Cosper at the University of Louisiana at Lafayette, and the Louisiana Board of Regents for support. S.C.W. would like to thank Eric M. Schlegel for useful discussions.

\end{document}